\documentclass[%
 reprint,
amsmath,amssymb,
aps,
prb,
]{revtex4-2}
\usepackage[T2A]{fontenc}

\usepackage[russian,english]{babel}
\usepackage{comment}

\usepackage{physics}
\usepackage{xcolor}
\usepackage{tcolorbox}
\usepackage{ textcomp }
\definecolor{urlcolor}{HTML}{3333B2}
\definecolor{citecolor}{HTML}{3333B2}
\definecolor{black}{HTML}{000000}

\usepackage[
    pdftex, 
    pdftitle={Article},
    pdfauthor={Author}
]{hyperref} 
\hypersetup{citecolor=citecolor, urlcolor=urlcolor, colorlinks=true}

\usepackage{mathtools}

\newcommand{\iu}{\mathrm{i}\mkern1mu}

\newcommand{\ve}[1]{\textbf{#1}}

\usepackage{multirow}
\usepackage{graphicx}
\usepackage{dcolumn}
\usepackage{bm}

\begin{document}
\title{
Exceptional points in single open acoustic resonator due to the symmetry breaking
}
\author{Vladimir Igoshin}
\author{Mariia Tsimokha}
\author{Anastasia Nikitina}
\author{Mihail Petrov}
\email{m.petrov@metalab.ifmo.ru}
\author{Ivan Toftul}
\author{Kristina Frizyuk}
\email{k.frizyuk@metalab.ifmo.ru}
\affiliation{The School of Physics and Engineering, ITMO University, Saint-Petersburg 197101, Russia}

\begin{abstract}

Exceptional points (EPs) have been widely studied in quantum mechanics, condensed matter physics, optics and photonics. However, their potential in acoustics has only recently been recognized due to the rapid development of acoustic metamaterials. This paper proposes a method for achieving EP conditions in acoustic resonators by lowering their symmetry and enabling resonant mode interaction. The formation of EPs is predicted through direct numerical simulation supported by coupled mode theory and resonant state expansion. These findings have significant implications for the design and optimization of acoustic metamaterials for applications such as acoustic sensing and noise reduction.

\end{abstract}


\maketitle

\section{\label{sec:introduction}Introduction}

Acoustic metamaterials are a promising class of materials that offer unique capabilities for tailoring properties of sound waves~\cite{Cummer2016Feb,Fok2008Oct,Ma2016Feb} as well as for mechanical manipulation~\cite{Toftul2019-AcousticRadiationFo,Shi2019Jul,Sepehrirahnama2022Oct}. 
While resonances play a central role in acoustics metamaterials there are still many hindered physical effects  and  mechanism.
EPs are the points in the parameter space where the eigenvalues of the system become degenerate and the eigenvectors coalesce, leading to a non-diagonalizable Jordan block formation~\cite{Kato, Heiss2012Oct, Schafer2022-Symmetryprotectedex, Moiseyev2011Feb}. The  spectral  singularities related to EP  are   highly sensitive to  the system parameters making them very prospective for sensing applications~\cite{Wiersig2020Sep, Park2020Apr}. 
Despite the progress in this area, the problem of EP appearance in single acoustic resonators still  requires thorough study and it is addressed in this paper.

EPs occur only in non-hermitian systems~\cite{Moiseyev2011Feb} and they are often mentioned in the context of PT-symmetric systems~\cite{AluEP, Heiss2012Oct, El-Ganainy2018Jan, Feng2017Dec} observed in  electronics~\cite{Schindler2011Oct},  optics and photonics~\cite{AluEP, Wiersig2020Sep, Feng2017Dec}, and recently in acoustics~\cite{Achilleos2017Apr, Fleury2015Jan}. However, reaching PT-symmetry requires particular engineering of gain and loss in acoustical systems. Alternatively, EP can be observed in open resonators as  a special class of non-Hermitian systems, which was extensively studied in optics and photonics~\cite{Feng2017Dec,Midya2018Jul,Pan2018Apr}. One of the possible mechanisms of EP formation is based on  \textit{breaking  the symmetry of the resonator}~\cite{Valero2022May}.
While this is not the unique approach we leave the other methods beyond the scope of the current work and refer the readers to Ref.~\onlinecite{Heiss2012Oct,Cao2015-Dielectricmicrocavit}.
However, despite extensive research on EPs in optics, there has been little work on EPs in the acoustics domain. 

\begin{figure}[ht]
\centering
\includegraphics{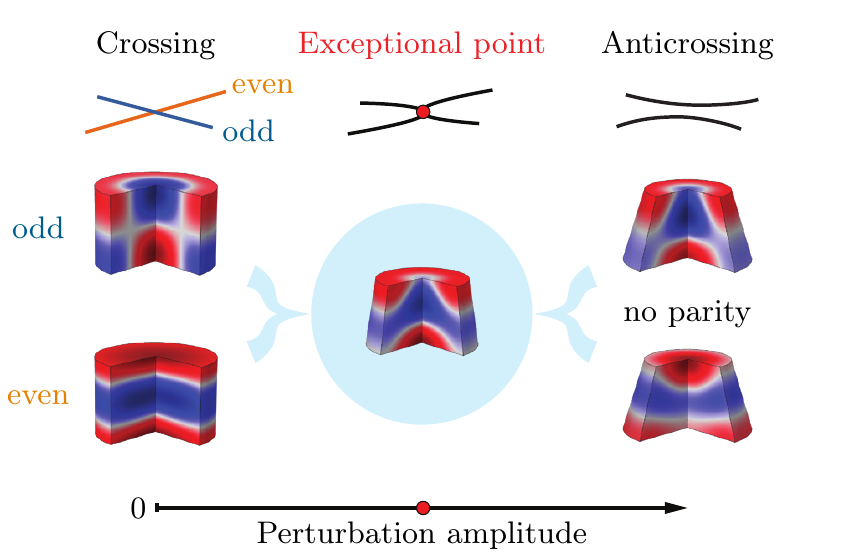}
\caption{The main idea of this work. 
Symmetry breaking perturbation merges two groups of modes induced by symmetry considerations into a single group.
This is shown at the bottom of the figure.
By changing amplitude of the symmetry breaking perturbation we can tune a coupling strength.
As shown in the top of the figure, this allows us to observe the transition from crossing of energy terms to avoided crossing.
This transition can be followed by EP characterized by the coalescence of eigenspaces.} 
\label{fig:concept}
\end{figure}

In this work, we show that by engineering the shape of resonators and due to related symmetry breaking one can enable  mode coupling mechanisms leading to formation of  EP  condition as schematically shown in Fig.~\ref{fig:concept}: in a system with perturbed symmetry two initially non-interacting modes transform into two different modes a degenerate  a state. This transitional state appears to be  an EP state.

We improve and adapt the powerful method of multipolar  analysis~\cite{SymmetryClass,SymmetryGladyshev,Sadrieva2019Sep} to \textit{(i)} predict the occurrence of EPs as a result of a particular symmetry breaking and \textit{(ii)} to understand deeper the mode interaction within a coupled mode theory and resonant state expansion (RSE) method.

Exploring the physics behind the EPs formation in acoustic resonators may unlock novel methods for the design and optimization of acoustic metamaterials for a wide range of applications such as sound focusing~\cite{Guenneau2007Nov}, optomechanics~\cite{Eichenfield2009Nov}, sensing~\cite{Nirschl2009Nov,Thomas2016Jun,Lee2020Mar,Moleron2015Aug}, noise insulation~\cite{Krasikova2023-Metahouse:Noise-Insu}, seismic cloaking~\cite{Muhammad2022Mar,Gupta2023Feb}.
By addressing the important questions surrounding the physics of acoustic metamaterials, we may open the door to new and exciting opportunities in acoustics and materials science. 

This work is organised as follows. 
{In Section~\ref{sec:coupled_modes} we build a simple model based on the linear acoustic equations
and discuss the simple mechanism of the EP appearance. 
In Section~\ref{sec:general} we extend  the approach to more complex symmetries and generalize the method. Finally, in Section~\ref{sec:conc} we conclude the obtained results.  }

\section{Coupling of modes in acoustic resonators}
\label{sec:coupled_modes}

\begin{figure}
	\includegraphics{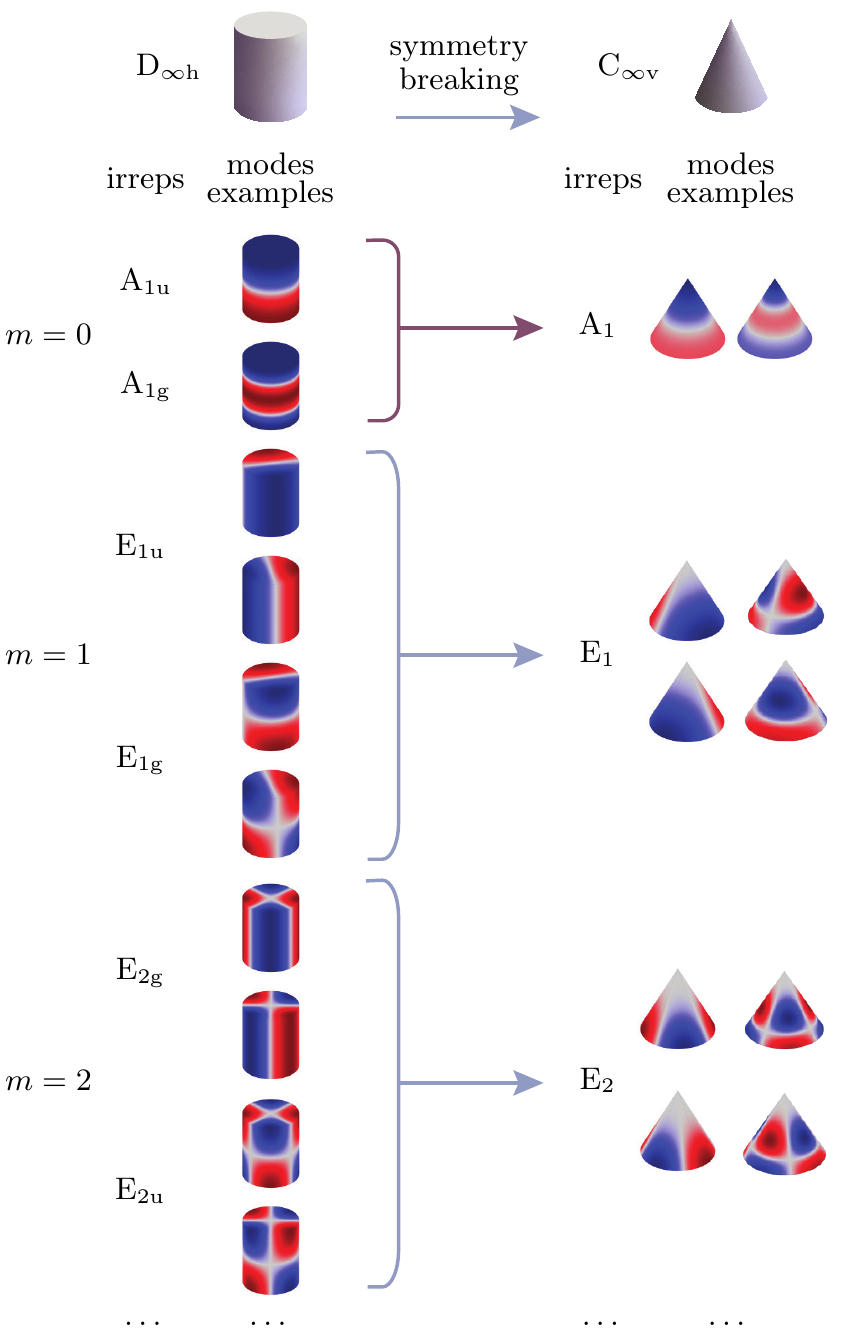}
	\caption{
 Tables of irreducible representations of symmetry groups $\text D_{\infty \text h}$ and $\text C_{\infty \text v}$ and examples of the eigenmodes, transforming under these irreps. Only the first few irreps are shown. Monopole mode transforms under irreducible representation  $\text{A}_{1\text{g}}$ but omitted in the figure in favour of a more illustrative mode.
 Examples of modes that transform under the particular irreducible representations are given.
 The arrows show how the irreducible representations of the two groups are related to each other.
	}
	\label{fig:IrrepsModes}
\end{figure}
\subsection{Model physics and parameters}

We focus on the studies of  resonators in the framework of linear acoustics  in the frequency domain. The resonators are made of homogeneous acoustics materials and placed in homogeneous environment (fluid or gas) which supports only longitudinal waves. Both media are characterized by their  compressibility $\beta$ and mass density $\rho$ (the speed of sound is $c = 1/\sqrt{\beta \rho}$). 
The complex velocity $\vb{v}(\vb{r})$ and pressure $p(\vb{r})$ fields satisfy the wave equations~\cite{LandauFluidM}
\begin{equation}\label{eq:acoustics}
    \iu \omega \beta p = \div \vb{v}, \qquad
    \iu \omega \rho \vb{v} = \grad p.
\end{equation}
By assuming the harmonic convention  $\exp(-\iu \omega t)$ and applying the Sommerfeld radiation condition, one can end up with a generalized eigenvalue problem
\begin{equation}
\label{eq:GEigV}
    \underbrace{\begin{pmatrix}
            0 & -\div\\
            \grad & 0 
    \end{pmatrix}}_{ \hat{\mathbb{D}}}
    \underbrace{\begin{pmatrix}
            p \\
        \iu \mathbf{v}
    \end{pmatrix}}_{\vec{\mathbb{F}}}
    =
    \omega
    \underbrace{\begin{pmatrix}
            \beta (\ve r) & 0\\
            0 & \rho (\ve r)
    \end{pmatrix}}_{\hat{\mathbb{P}}}
    \underbrace{
    \begin{pmatrix}
            p \\
            \iu \mathbf{v}
    \end{pmatrix}}_{\vec{\mathbb{F}}},
\end{equation}
which describes eigenmodes $\vec{\mathbb{F}}(\mathbf{r})$ of an open acoustic resonator, where operators $\hat{\mathbb{D}}(\mathbf{r})$ and $\hat{\mathbb{P}}(\ve r)$ are introduced for future convenience. 
The formulated eigenvalue problem is in a full analogy to the one appearing in optics~\cite{Muljarov2018May}. 


Eigenmodes of this system can be classified by irreducible representations (irreps) of a symmetry group of a resonator~\cite{Zee2016Mar,Landau1977,SymmetryClass} similarly to the case of optical resonators~\cite{SymmetryGladyshev}. 
Some particular material parameters have to be chosen to illustrate general conclusions of this paper which, in general, do not depend on this choice.
For the most acoustic experiments performed in air, the wavelength in the scatterer is longer than in the host medium. However, we have chosen our parameters in favour of numerical stability since all conclusions do not depend on whether acoustic refractive index $\sqrt{{\beta_1\rho_1}/{\beta_0\rho_0}}$ is greater or less then one.
To be specific, we use density $\rho_0 = 1~\text{kg}/\text{m}^3$ and speed of sound $c_0 = 1~\text{m/s}$ for the media and $\rho_1 = \rho_0/2$, $c_1 = c_0/2$ for the resonator.
Needless to say that similar scenario is also possible in practice: high-index resonator material is typical for Mie-resonance acoustic meta-atoms, which can be used to create metamaterials~\cite{Cheng2015Oct,Shi2019Jul,Liang2012Mar,Melnikov2019Jul}.
Nevertheless, we stress again that the theory presented below does not depend on the material parameters of the media and the resonator.

The eigenmodes computations were performed with help of numerical simulations in COMSOL Multiphysics\textsuperscript{\textregistered} and the method described briefly in Appendix~\ref{sec:COMSOL}.

\subsection{EP appearance}

\begin{figure}
    \includegraphics{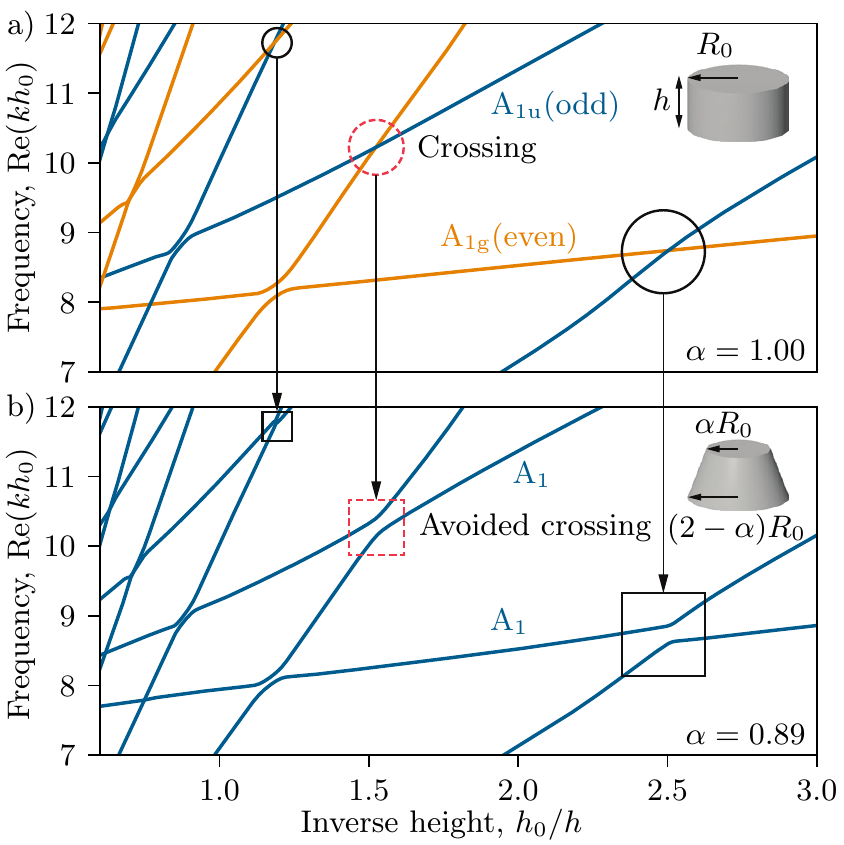}
    \caption{Real parts of eigenvalues $k$ scaled to $h_0$ of a cylinder with $\alpha=1$ (a) and a conical frustum with $\alpha=0.89$ (b) versus $h_0/h$. The branches of cylinder's modes have different colors according to their irrep. Intersections of modes that transform under different irreps are circled. Corresponding avoided crossings and crossing with weak coupling are marked with a square. The crossing and the avoided crossing shown in Fig.~\ref{fig:EPcomsol} are dashed. Geometry parameters used are $h_0=1\text{ m}$, $R_0=0.5\text{ m}$.}
    \label{fig:modes_cone_to_cyl}
\end{figure}

In order to clarify the connection between the symmetry and EP formation, we firstly will elaborate on  the resonator spectrum  modification with change of the resonator shape. We start with a structures of high symmetry such as cylinders~\cite{Tsimokha2022Apr} and move to more complex shapes later. We fix the initial geometry of the cylinder such that height to radius ratio is $h_0/R_0=2$ [see Fig.~\ref{fig:modes_cone_to_cyl}  (inset)].


Cylinder corresponds to $\text D_{\infty \text h}$ symmetry group, thus it has an countably infinite set of finite-dimensional irreducible representations~\cite{Landau1977-QuantumMechanics, BibEntry2021Mar,BibEntry2001Sep} as shown in Fig.~\ref{fig:IrrepsModes}, where each representation is demonstrated along with corresponding examples of numerically calculated pressure fields. 
Each of those irreducible representations of a symmetry group of a cylinder corresponds to a single value of the azimuthal number $m$.  
In the following, we narrow the consideration to $m = 0$ only, or, in more specific terms, eigenmodes which transformed under irreducible representations A$_\text{1u}$ and A$_\text{1g}$. 
These eigenmodes have rotational symmetry, but different parity under horizontal reflection ($\sigma_{\text{h}}$ transformation). The obtained results can be straightforwardly extended to other azimuthal numbers $m$. We have omitted the monopole mode which transforms under irreducible representation  $\text{A}_{1\text{g}}$ in favour of another more illustrative example. 
However, we stress that during the symmetry braking $\text{D}_{\infty \text{h}} \to \text{C}_{\infty \text{v}}$ (e.g. cylinder to cone), monopole and dipole modes start to transform under the same irreducible representation $\text{A}_{1}$ in the cone symmetry group and hence interact with each other. 
Resonators which exhibit this feature are conventionally characterised by Willis coupling~\cite{Sepehrirahnama2022Oct,Quan2018Jun,Sieck2017Sep,Willis1985Jan}.

The dependence of the real part of eigenfrequency of the $m=0$ modes on the resonator's height is shown in  Fig.~\ref{fig:modes_cone_to_cyl}~(a), where each color of the line corresponds to its own irreducible representation A$_\text{1u}$ or A$_\text{1g}$. {Hereafter, we would call eigenmodes transformed under irreducible representations A$_\text{1u}$ and A$_\text{1g}$ as mode A$_\text{1u}$ and mode A$_\text{1g}$ respectively.} At this point,  it should be mentioned that modes transforming under different irreps are orthogonal, therefore coupling between them does not occur~\cite{Zee2016Mar, Landau1977}. In Fig.~\ref{fig:modes_cone_to_cyl}~(a), this statement can be interpreted visually, where the crossing between two orthogonal modes of a cylinder, A$_\text{1u}$ and A$_\text{1g}$, occurs. Indeed, explicit analysis of the  eigenmodes' fields show that these modes have different parities, 'odd' and 'even' correspondingly,  with respect to $\sigma_{\text{h}}$ (reflection in the horizontal plane) transformation. On the other hand, Fig.~\ref{fig:modes_cone_to_cyl}~(a) shows the  avoided crossing behavior of the energy lines of the mode related to the same irrep.

Next, one can alter  the symmetry of the resonator by perturbing its  shape from cylinder to cone, which breaks the mirror symmetry in horizontal plane. In order to carefully trace out changes in the mode structure those symmetrical changes should be made gradually, so we chose the  parameter of  ``cylindricity'' of the resonator's shape $\alpha$  Fig.~\ref{fig:modes_cone_to_cyl}~(b),  which denotes the amplitude of the shape perturbation. 
Varying  $\alpha$ from 1 to 0 reduces the top radius of the cylinder, while simultaneously increasing the bottom radius.
Thus, decreasing the $\alpha$  parameter breaks the symmetry with respect to $\sigma_{\text{h}}$ and changes the symmetry group of the system from $\text D_{\infty \text h}$ to $\text C_{\infty \text v}$. According to Fig.~\ref{fig:IrrepsModes} previously considered A$_\text{1u}$ and A$_\text{1g}$ modes lose their specific $\sigma_{\text{h}}$ parity property, and now transform under the same irrep A$_\text{1}$. Therefore, a coupling between those modes should now appear in conical symmetry which manifests itself in appearance of avoided crossings between the frequency lines shown in Fig. \ref{fig:modes_cone_to_cyl}~(b). The transition from crossing to avoided crossing induced by symmetry breaking is shown with arrows.   

\begin{figure*}[t]
    \includegraphics{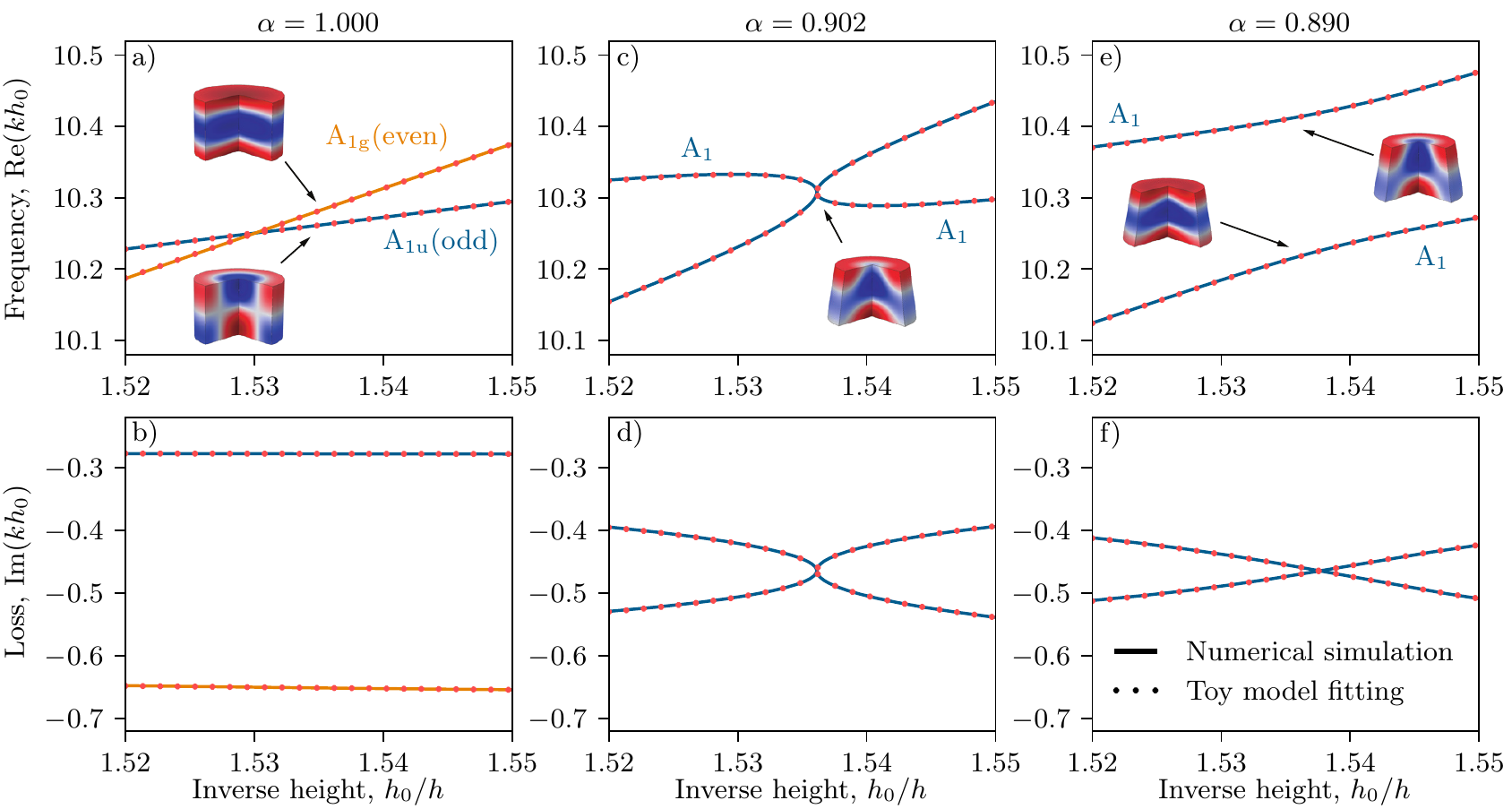}
    \caption{Maps of real and imaginary parts of eigenvalues $k$ of the acoustic resonator (scaled to $h_0$). Examples of fields in resonators are given.
    (a) The crossing of real parts of two non-interacting modes that transform under different irreps $\text{A}_\text{1u}$ and $\text{A}_\text{1g}$ is observed at ``cylindricity'' parameter $\alpha = 1.000$.
    (e) At $\alpha = 0.890$ there is an anticrossing of real parts of modes that transform under the same irrep $\text{A}_\text{1}$ (strong coupling).
    At $\alpha = 0.902$ an EP is achieved in which both real (c) and imaginary (d) parts of eigenvalues are crossed, and therefore modes degenerate.
    A toy model fitting is discussed in Appendix~\ref{sec:toy_model_fitting}.}
    \label{fig:EPcomsol}
\end{figure*}

To this point, almost exclusively the real part of the eigenfrequency was discussed. Imaginary parts of the   eigenfrequencies are shown in Fig.~\ref{fig:EPcomsol}~(b). While a crossing behaviour in real parts [Fig.~\ref{fig:EPcomsol}~(a)] can be observed, imaginary parts are not equal, and the modes are not degenerate.
However, by varying the height of the resonator $h$ and  the cylindricity parameter $\alpha$, at a certain coupling strength, the simultaneous degeneracy  of real and imaginary parts of two modes is observed [see  Fig.~\ref{fig:EPcomsol}~(c,d)]. Now the series of 
Fig.~\ref{fig:EPcomsol}~(a, c, e) can be regarded as follows: from left to right three different stages of the resonator's deformation from a cylinder to a cone are shown. At the most left (a,b), a $\text D_{\infty \text h}$ symmetry of the resonator ensures a complete absence of any interaction between the two considered modes; at the most  right (e,f) the interaction is great enough to keep those modes entirely apart from each other; evidently, based on the smoothness of this particular symmetry breaking, at the intermediate parameters shown in (c,d), a point corresponding to full degeneracy of two those two modes can be found. In Fig.~\ref{fig:EPcomsol}~(c, d) two lines merge into a single point, both for the real and imaginary parts of the frequency, which corresponds to formation of the EP. Note, that  usual degeneracy of two modes happens in so-called diabolic points \cite{Moiseyev2011Feb}, which correspond to the two- or more dimensional irreps, but here the fact that A$_\text{1}$ irrep is one-dimensional make this even not possible.
{In simple words, we cannot obtain two modes with the same frequency whose field are linearly independent~\cite{AluEP, RevModPhys.87.61, Wiersig2016Mar, Wiersig2020Sep}}. One should also note, that the crossing-to-avoided crossing  mechanism is precisely inverted for real and imaginary parts, as shown in  Fig.~\ref{fig:EPcomsol}~(b, d, f), where the system goes from an anti-crossing, through a degeneracy, to a crossing.
The observed modes interaction via symmetry breaking can be extended to other shapes of the resonators as discussed in Appendix~\ref{sec:c4v}.


\subsection{Coupled mode theory and perturbation of the material parameters}
\label{sec:toy_model}

The appearance of EP in quantum mechanics and optics can be often interpreted within a scope of a standard coupled mode theory, described by a 2$\cross$2 Hamiltonian of a two-level system~\cite{AluEP, RevModPhys.87.61, Wiersig2016Mar, Wiersig2020Sep, Heiss2012Oct}.
While coupling of several acoustic resonators has already been analysed withing standard coupled mode theory \cite{Maksimov2015-Coupledmodetheoryf,Miroliubov2021-SpectralCharacterist}, the mode coupling in a single acoustic resonator upon its shape perturbation has not yet been discussed. In this section, we introduce the phenomenological model of interaction of two eigenmodes based on resonant-state expansion (RSE) or quasinormal mode expansion~\cite{Doost2014RSE, LalanneQNMRev, Muljarov2018May, LalanneMAN}.  
RSE has already proved its efficiency for determining the eigenmodes  of perturbed optical system, appearance of strong coupling regime~\cite{BogdanovBIC}, and multipole coupling~\cite{SymmetryGladyshev}.
However, as far as we know, no such analysis has yet been provided for linear acoustics described by Eq.~\eqref{eq:GEigV}.
This equations can be rewritten in a compact form similar to Ref.~\onlinecite{Muljarov2018May}:
\begin{equation}
    \hat{\mathbb{D}}(\mathbf{r})\vec{\mathbb{F}}_n(\mathbf{r}) = \omega_n \hat{\mathbb{P}}(\mathbf{r})\vec{\mathbb{F}}_n(\mathbf{r}) 
\end{equation}
where  $\vec{\mathbb{F}}_n(\mathbf{r}) = \begin{pmatrix}p_n, & \iu \mathbf{v}_n\end{pmatrix}^\intercal$ is an eigenmode, $\omega_n$ is an eigenvalue.
For the perturbed resonator there is a new set of eigenmodes $\vec{\mathbb{F}}(\mathbf{r})$ and eigenvalues $\Omega$
\begin{equation}
\hat{\mathbb{D}}(\mathbf{r})\vec{\mathbb{F}}(\mathbf{r}) = \Omega \left(\hat{\mathbb{P}}(\mathbf{r}) + \Delta\hat{\mathbb{P}}(\mathbf{r})\right)\vec{\mathbb{F}}(\mathbf{r})     
\end{equation}
where $\Delta\hat{\mathbb{P}}(\mathbf{r})$ is the perturbation of the material parameters, caused, for example, by changing the resonator's shape.

It is assumed that all eigenmodes are a linear combination of eigenmodes of the unperturbed system $\vec{\mathbb{F}}(\mathbf{r}) = \sum\limits_n c_n \vec{\mathbb{F}}_n(\mathbf{r})$.
Within this approach, the  perturbation theory formalism~\cite{Muljarov2018May} immediately provides us with the coupled equations system
\begin{equation}
    \sum_n \left((\Omega-\omega_n)\delta_{mn} + \Omega V_{mn}\right)c_n = 0
    \label{eq:matrix_eq}
\end{equation}
where $V_{mn} \propto  \displaystyle \int \vec{\mathbb{F}}_m^\intercal \Delta\hat{\mathbb{P}}\vec{\mathbb{F}}_n$ is the matrix element. 
In the two-mode approximation of Eq.~\eqref{eq:matrix_eq}, we obtain the eigenvalue problem
\begin{equation}
    \begin{pmatrix}
            \omega_1 & 0\\
            0 & \omega_2
    \end{pmatrix}
    \begin{pmatrix}
            c_1\\
            c_2
    \end{pmatrix}=
    \Omega
    \begin{pmatrix}
            1 + V_{11} & V_{12}\\
            V_{12} & 1 + V_{22}
    \end{pmatrix}
    \begin{pmatrix}
            c_1\\
            c_2
    \end{pmatrix}
\label{twomodeapprox}
\end{equation}
The eigenvalues of this equation are
\begin{equation}
\label{eq:CMT_freq}
    \Omega_\pm = \frac{\xi\omega_2+\gamma\omega_1\pm\sqrt{(\xi\omega_2-\gamma\omega_1)^2+4\kappa^2\omega_1\omega_2}}{2(\xi\gamma-\kappa^2)}
\end{equation}
where $\kappa =V_{12} = V_{21}$, $\xi = 1+ V_{11}$, $\gamma = 1+V_{22}$.
It can be seen that the non-diagonal element $\kappa$ is responsible for the coupling between the modes.

The main condition of observation of EP can be summarized as $\Omega_- = \Omega_+$. Or, in a more extended form $(\xi\omega_2-\gamma\omega_1)^2+4\kappa^2\omega_1\omega_2=0$.
It can be seen that if this condition is satisfied and $\kappa\neq0$, the eigenspaces collapse.

Here, we show that this system can be applied to describing the EP formation in acoustic resonators with symmetry breaking. The exact identification of governing parameters $\kappa$, $\xi$, and $\gamma$ requires  introducing the proper scalar product and normalization and stays beyond the scope of the current work. Still, we found these parameters numerically by fitting the real and imaginary frequency curves in Fig.~\ref{fig:EPcomsol} within Eq.~\eqref{eq:CMT_freq} (see Appendix \ref{sec:toy_model_fitting} for the details).  The obtained results are shown with dots in Fig.~\ref{fig:EPcomsol} demonstrating excellent correspondence to the  numerical results.

Within the coupled mode approach  it becomes evident that  two modes that transform under the same irrep may couple to each other since the coupling constant provided by the integral $\kappa \propto  \displaystyle \int \vec{\mathbb{F}}_1^\intercal \Delta\hat{\mathbb{P}}\vec{\mathbb{F}}_2 \neq 0$ in full accordance with Wigner's theorem~\cite{Solyom2007-ConsequencesofSymme}. It can also be derived from the  selection rules presented in~\cite{Zee2016Mar, Landau1977}.

\section{Generalization of the approach to other symmetries}
\label{sec:general}

The illustrated mechanism of EP formation can be extended to resonators of arbitrary shapes. The general recipe for that is as follows: 

\begin{enumerate}
    \item Two non-interacting eigenmodes, which transform under different irreps have to be chosen.
    \item Next, by particular  symmetry breaking  one needs to assure that these modes will appear in the same irrep. This step is essential, as the chosen symmetry will trigger the formation of EP states.
    \item Now the original symmetry has to be broken smoothly for transition from an object of the original symmetry, to one of the chosen new symmetry. In other words, we should have a homotopy equivalence between two shapes. Thus, the shape change is moderated by one parameter.  
   This way, an appearance of EP can be traced easily.
\end{enumerate}

Following this  algorithm, one can observe EP formation in more complex shapes and symmetries of the resonator.   As an example, besides already  considered $\text D_{\infty \text h}$ to $\text C_{\infty \text v}$ symmetry breaking, one can observe EP formation in $C_{4v}$  to $C_{2v}$ symmetry breaking as discussed in details in Appendix~\ref{sec:c4v} and supported by numerical simulations. Formation of EP is provided, for example by the merging of $B_1$ and $A_1$ irreps into $A_1$.

%


\section{Conclusion}
\label{sec:conc}
In this work, a general mechanism of retrieving exceptional points in complex acoustic systems is proposed. We identify an approach to  achieving exceptional point condition by breaking a particular symmetry of the resonator leading to coupling of initially orthogonal eigenmodes. Smooth variation of asymmetry parameter allows for finding the exact conditions for mode coalescence. This mechanism was illustrated in single homogeneous acoustic resonators by breaking their symmetry from $\text D_{\infty \text h}$ to $\text C_{\infty \text v}$, from $C_{4 \text v}$  to $C_{2 \text v}$. 
The results were supported by direct numerical simulations. Moreover, we proposed a phenomenological model of interaction of two eigenmodes based on resonant state expansion method and  perturbation theory.
We have verified that the proposed description in the two-mode approximation can be used for qualitative analysis of eigenfrequencies.

\begin{acknowledgments}
The numerical and analytical simulations were supported by Russian Science Foundation grant no. 20-72-10141. The work is financially supported by the Program Priority 2030. MP acknowledges support from BASIS Foundation.
\end{acknowledgments}


\appendix

\section{Example of $\text{C}_{4\text{v}} \to \text{C}_{2\text{v}}$ transition}
\begin{figure}[ht]
\includegraphics{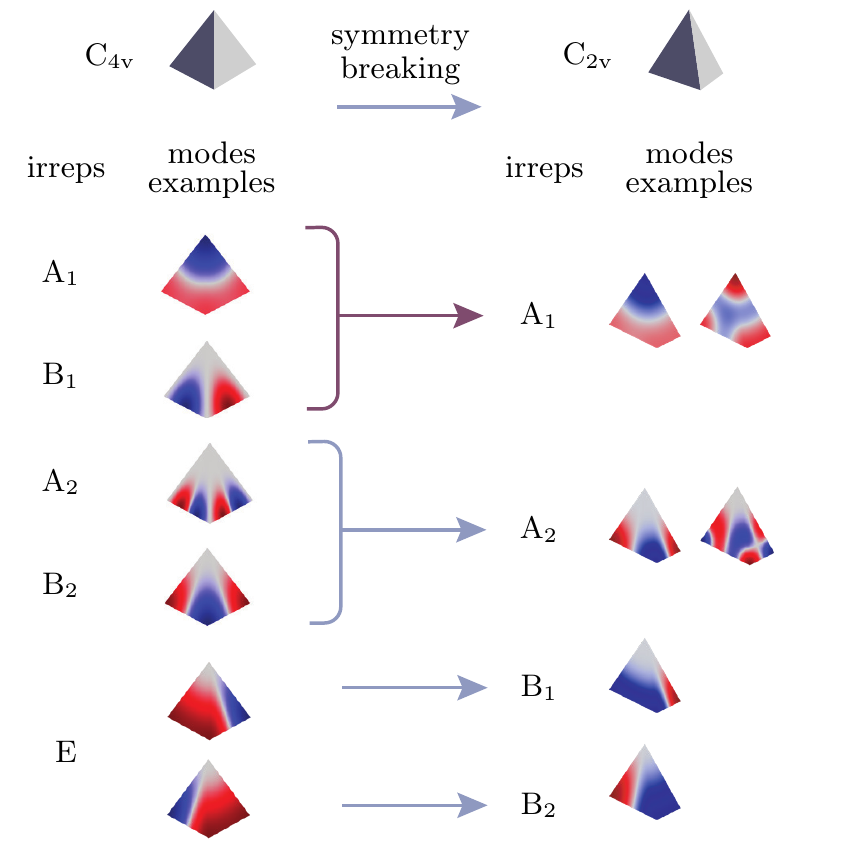}
\caption{
Tables of irreducible representations of symmetry groups $\text C_{4 \text v}$ and $\text C_{2 \text v}$.
Examples of modes transformed under particular irreducible representations are given.
The arrows show how the irreducible representations of the two groups are related to each other. 
}
\label{fig:c4v}
\end{figure}

\label{sec:c4v}

\begin{figure}[ht]
    \includegraphics{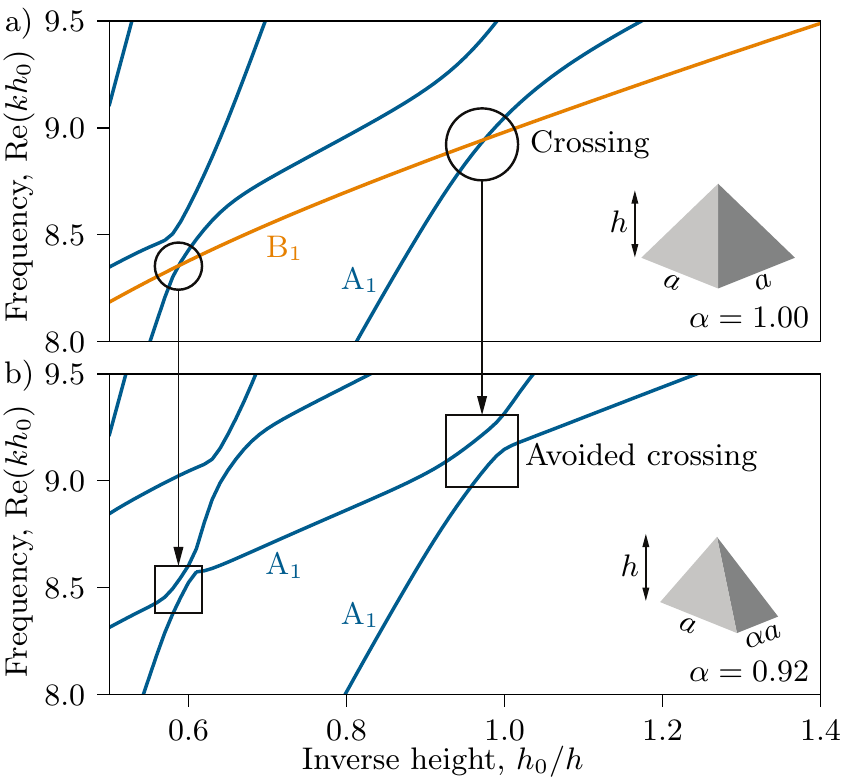}
    \caption{Real parts of scaled with $h_0$ eigenvalues $k$ of a square right pyramid ($\alpha=1$) and a rectangular right pyramid ($\alpha=0.92$) versus $h_0/h$.
    The branches of the square pyramid modes have different colors according to their irrep.
    Intersections of modes that transform under different irreps are circled. The corresponding avoided crossings are marked by a square.
    Geometry parameters used are $h_0=1\text{ m}$, $a=1\text{ m}$.
    }
    \label{fig:modes_c4v_to_c2v}
\end{figure}
{The mechanism of EP formation described above is very general and is based on symmetry requirements.}

{Having two modes which show a crossing behaviour in the unperturbed system, we  break the symmetry in the way that makes this modes to interact (i.e. merges two irreps into a single one, so two modes, which transform by the different irreps will be transformed by the same irrep after the perturbation).}
{The EP state occurrence for $\text D_{\infty \text h}\to \text C_{\infty \text v}$ symmetry breaking, which has already been considered above, can be generalized to more complex cases such as a transition $\text C_{4\text v}\to \text C_{2 \text v}$.

The merging $\text A_1$ and $\text B_1$ modes of regular square pyramid $\text C_{4\text v}$ to $\text A_1$ modes of rectangular right pyramid $\text C_{2 \text v}$ is shown in Fig.~\ref{fig:c4v}, while Fig.~\ref{fig:modes_c4v_to_c2v} illustrates crossing between the two orthogonal modes $\text A_1$ and $\text B_1$ of the regular square pyramid and avoided crossing between the two modes of the rectangular right pyramid that transform under same irreducible representation $\text A_1$. In this case avoided crossing and exceptional points can be easily achieved due to the symmetry breaking and the proper selection of parameters.}

\section{Toy model fitting}
\label{sec:toy_model_fitting}
In order to test the ability of the model (Section~\ref{sec:toy_model}) to describe the effects presented in the Fig.~\ref{fig:EPcomsol}, we fit its parameters $\xi$, $\gamma$, $\kappa$.
A range of heights $h_0/h$ from $1.52$ to $1.55$ is used, model parameters are assumed to be constant. Eigenfrequencies for unperturbed system are approximated as 
\begin{align*}
    \omega_1(h_0/h)=(2.224724-\iu 0.023396 )\cdot  h_0/h \\
    +6.846411- \iu 0.242355
\end{align*}
and
\begin{align*}
    \omega_2(h_0/h)=(6.302923-\iu 0.221363)\cdot h_0/h \\
    +0.606508- \iu 0.31179
\end{align*}

The results of fitting for each $\alpha$ value are presented in Table~\ref{tab:toymodel}.
We choose model parameters for which there is a sufficiently accurate match with the numerical experiment.

\begin{table*}
\caption{\label{tab:toymodel} Model parameters for different $\alpha$ values and their corresponding fitting relative errors.}
\begin{ruledtabular}
\begin{tabular}{cccccccc}
    $\alpha$ & $\xi$ & $\gamma$ & $\kappa$ & max rel. error, \%\\ \hline
    $1.000$  & $1$ & $1$ & $0$ & $0.0044$ \\ 
    $0.902$ & $0.993783+ \iu 0.000728$ & $1.000755- \iu 0.001037$ & $0.017148- \iu 0.001479$ & $0.0534$\\ 
    $0.890$ & $0.992245+ \iu 0.000903 $ & $1.000909- \iu 0.001272 $ & $0.019272- \iu 0.001574$ & $0.0113$\\ 
\end{tabular}
\end{ruledtabular}
\end{table*}

\section{Numerical modelling}
\label{sec:COMSOL}
All simulations were performed using the Pressure Acoustics branch of Acoustics module of COMSOL Multiphysics\textsuperscript{\textregistered} software.
The simulation domain was a sphere made of media material with the resonator inside.
When searching eigenmodes for Fig.~\ref{fig:EPcomsol}, the perfectly matched layer (PML) was used to simulate open boundary conditions on the surface of the spherical domain.
To calculate eigenmodes for Fig.~\ref{fig:modes_cone_to_cyl} and Fig.~\ref{fig:modes_c4v_to_c2v} the radiation boundary condition for a spherical wave was used instead of the PML in order to speed up calculations.
All numerical modes (also termed as PML modes) were removed during post-processing~\cite{LalanneQNMRev}.

\bibliography{mode_coupling_refs}

\end{document}